\documentclass[aps,preprint]{revtex4}
\usepackage{psfig}
\usepackage{amsmath}
\usepackage{amssymb}

\begin{document}

\title{Design of a folding inhibitor of the HIV--1 Protease}
\author{R. A. Broglia$^{1,2,3}$, G. Tiana$^{1,2}$, D. Provasi$^{1,2}$, F. Simona$^{1,2}$, L. Sutto$^{1,2}$, F. Vasile$^4$ and M. Zanotti$^{1,2}$}
\address{$^{1}$Dipartimento di Fisica, Universit\'a di Milano, via Celoria 16, 20133 Milano, Italy}
\address{$^{2}$INFN, Sez. di Milano, Milano, Italy}
\address{$^{3}$Niels Bohr Institute, University of Copenhagen, Bledgamsvej 17, 2100 Copenhagen, Denmark}
\address{$^{4}$Dipartimento di Scienze Molecolari Agroalimentari, University of Milano, via Celoria 2, Milano, Italy}

\begin{abstract}
Being HIV--1--PR an essential enzyme in the viral life cycle, its inhibition can control AIDS. Because the folding of single domain proteins, like HIV--1--PR is controlled by local elementary structures (LES, folding units stabilized by strongly interacting, highly conserved amino acids) which have evolved over myriads of generations to recognize and strongly attract each other so as to make the protein fold fast, we suggest a novel type of HIV--1--PR inhibitors which interfere with the folding of the protein: short peptides displaying the same amino acid sequence of that of LES. Theoretical and experimental evidence for the specificity and efficiency of such inhibitors are presented.
\end{abstract}
\date{\today}
\maketitle

\section{Introduction}

HIV--1--PR is a homodimer, that is a protein whose native conformation is built of two (identical) disjoint chains each of them made of 99 amino acids. Sedimentation equilibrium experiments have shown that at neutral pH the protease folds according to a three--state mechanism ($2U\rightarrow 2N\rightarrow N_2$), populating consistently the monomeric native conformation $N$ \cite{xie}. The dimer dissociation constant ($2N\rightarrow N_2$) is $k_D=5.8\;\mu M$ at room temperature, while the folding temperature of the monomer, i.e. the temperature at which the free energy of the native monomeric state $N$ is equal to that of the unfolded state $U$ is $T_f=52.5$C \cite{rozzelle}.

Monte Carlo simulations of the folding of three--state homodimers \cite{dimer} indicate that after the monomers have reached the native state $N$, they diffuse to find another folded monomer to associate with. The same study also shows that in the stage $U\rightarrow N$, each monomer folds according to the same mechanism controlling the folding of single domain proteins. This mechanism is reasonably well understood as a result, among other things, of extensive lattice model studies as well as of all--atom off--latice G\=o--model simulations (cf. ref. \cite{varenna} and refs. therein). These studies revealed the central role played by few, strongly interacting highly conserved, as a rule hydrophobic, amino acids. Mutations of these "hot" amino acids lead, in general, to denaturation \cite{tiana98}. It has been shown \cite{hierarchy,jchemphys3} that, starting from an unfolded conformation, folding proceeds following a hierarchical succession of events: 1) formation of LES  stabilized by the interaction among "hot" amino acids lying close along the polypeptide chain, 2) docking of the LES into the (post critical) folding nucleus, that is formation of the minimum set of local contacts which brings the system over the major free energy barrier of the whole folding process, 3) relaxation of the remaining amino acids on the native structure shortly after the formation of the folding nucleus. Strong support for the soundness of this hierarchical scenario is found in a number of circumstantial evidences \cite{lesk,srinivasan,wallqvist,panchenko,hansen,rose,maritan,nussinov}.

The same model \cite{hierarchy,jchemphys3} suggests that it is possible to destabilize the native conformation of a protein with the help of peptides whose sequences are identical to those of the LES of the protein \cite{rudi}. Such peptides (p--LES) interact with the protein (in particular with their complementary LES) with the same energy which stabilizes its folding nucleus, thus competing with its formation.

One can mention two advantages of these non--conventional (folding--) inhibitors as compared to conventional (active site centerd--) inhibitors: I) their molecular structure is direclty fixed by that of the target protein. One has thus not to design a molecule so as to optimize its interaction with a given site of the enzyme, but just find which are the LES of the protein to be inhibited. The design of the p--LES having been performed by evolution. In fact, like the pair enzyme/substrate, pairs  of complementary LES have evolved through millions of years in order to learn how to recognize its partners as well as to avoid aggregation, II) it is unlikely that the protein can develop drug--resistance through mutations as LES are stabilized by hot amino acids.

The present paper is divided in two main parts. A theoretical one (Sections 2 and 3) where the prediction of the model are presented, and an experimental one (Sect. 4), where the properties of the designed inhibitor are tested.

\section{Folding units of HIV--1--PR: Local Elementary Structures}

The first step to be taken in the design of an inhibitor of the HIV--1--PR following the strategy discussed in the previous Section is to determine the LES of the target protein. In other words, segments of the protein containing highly conserved residues. To this scope use is made of evolution data input as well as model simulations. 

\subsection{Evolution}

A measure of the degree of conservation of residues in a family of proteins is the entropy per site $S(i)=-\sum_\sigma p_i(\sigma)\ln p_i(\sigma)$, where $p_i(\sigma)$ is the frequency of appearence of residue of type $\sigma$ at site i. Making use of a family of 28 uncorrelated proteins (i.e., displaying sequence similarity lower than 25\%) structurally similar to HIV--1--PR \cite{native}, the entropy $S(i)$ was calculated for all 99 sites of the monomers. From the result shown in Fig. \ref{fig1}(a) (continous curve) it is observed that the most conserved regions of the proteins involve residues 24-33, 56-61 and 81-87. Note that the conservation of residues 25-27 is not unexpected in that they build the active site of the protease.

Another important source of information concerning protected sites of the HIV--1--PR stems from drug-induced mutations observed in vivo.
The large production of virion in the cell, coupled with the error prone reverse transcriptase enzyme, eventually leads to escape mutants. Mutations observed in the stretches of the HIV--1--PR mentioned above are, as a rule, conservative mutations (e.g. L24I) \cite{tomasselli,wallqvist}. Less conservative mutations, e.g. hydrophobic to hydrophilic, are absent from these structurally protected regions.

\subsection{Model calculations: energetics} \label{sect_energy}

\begin{figure}
\centerline{\psfig{file=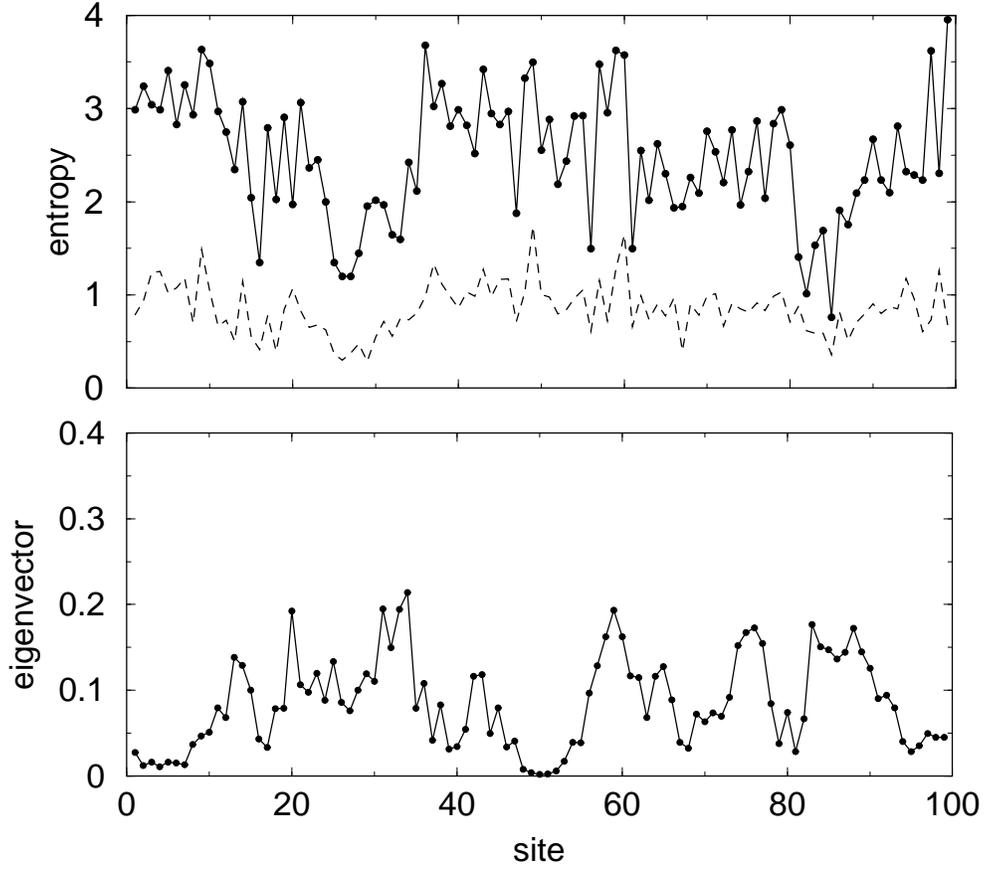,width=13cm,angle=-90}}
\caption{a) The entropy per site of proteins structurally similar to the HIV--1--PR monomer (pdb code: 1BVG). The solid line indicates the entropy function calculated over 28 proteins displaying sequence similarity lower than 25\% with the HIV--1--PR, while the dashed line is associated with 462 proteins irrespective of sequence similarity. b) The components of the eigenvector associated with the lowest eigenvalue of the interaction matrix between amino acids \protect\cite{giorgio}.}
\label{fig1}
\end{figure}

In order to become stable at an early stage of the folding process, LES must carry a significant fraction of the total energy of the protein. We have calculated this quantity making use of short all-atom molecular dynamics simulations in explicit water using GROMACS software \cite{gromacs}. The calculation was performed at 295 K for 1 ns, during which the protein fluctuates around its native conformation (the maximum RMSD being 11$\AA$). We have followed the scheme described in ref. \cite{giorgio}, calculating the average interaction energy $B_{ij}$ between any pair of amino acids during the 1ns dynamics and extracting the eigenvalues $\lambda_n$ of the resulting matrix $\parallel B_{ij}\parallel$. The lowest energy state ($\lambda_1$=-121.2 kJ/mol) displays a large energy gap (i.e. -12.9 kJ/mol$\approx 5kT$) with respect to the next eigenvalue, indicating a core of strongly interacting amino acids \cite{giorgio}. We display in Fig. \ref{fig1}(b) the eigenvector associated with the lowest energy eigenvalue, which highlights to which extent the different amino acids participate to this core. The largest amplitudes involve residues 25--32, 57--65, 74--77 and 83--90. Regions 25--32, 57--65 and 83--90 overlap well with the conserved regions mentioned above in connection with Fig. \ref{fig1}(a).

Another way of representing the interaction matrix $\parallel B_{ij}\parallel$ is to consider the interaction between fragments $S_1=(13-21)$, $S_2=(24-34)$, $S_3=(38-48)$, $S_4=(50-55)$, $S_5=(56-66)$, $S_6=(67-72)$, $S_7=(74-78)$, $S_8=(84-94)$. As can be seen from Table \ref{table1}, the corresponding $8\times 8$ energy map is essentially codiagonal, the associated energy of the interacting chain segment being $-1751$ kJ/mol as compared to the MD simulation native energy $-2533$ kJ/mol. In other words, the $S_1-S_8$ representation of the folded monomer accounts for $\approx 70\%$ of the calculated native conformation energy. In particular, the interaction between fragments $S_2$, $S_5$, $S_7$ and $S_8$ is stabilized by $-796$ kJ/mol equal to 32\% of the native conformation energy. Of this energy 20\% ($=-503$ kJ/mol) corresponds to the internal energy of the four fragments, while 12\% ($=-293$ kJ/mol) corresponds to the summed interaction energy between fragments.

 In keeping with the above results the $S_2$, $S_5$, $S_7$ and $S_8$ segments qualify as LES (folding units) of each of the two monomers of the HIV-1-PR dimer, LES which form in their native conformation the (post critical) folding nucleus (FN). It is interesting to note that drug induced mutations in the amino acids belonging to these LES (L24I, D30N, L33F, L63V, I64V, V77I, V82A, I84D, N88D and L90M \cite{tomasselli,wlodawer}), lead to a folding nucleus energy equal to -1500 kJ/mol, as compared to -796 kJ/mol for the wild type sequence FN, that is to increase of almost a factor of 2 in the stability of the system.

\subsection{Further evidence}

Wallqvist and coworkers\cite{wallqvist} investigated the HIV--1--PR molecule for the occurence of cooperative folding units that exhibit a relatively stronger protection against unfolding than other parts of the molecule. Unfolding penalities are calculated forming all possible combinations of interactions between segments of the native conformation and making use of a knowledge--based potential. This procedure identifies a folding core in HIV--1--PR comprising residues 22--32, 74--78 and 84--91, residues that form a spatially close unit of a helix (84--91) with sheet (74--78) above another $\beta$--strand ((22--25), containing the active site residues D25, T26 and G27) perpendicular to these elements.

Making use of a Gaussian network model, Bahar and coworkers \cite{bahar} have studied the normal modes about the native conformation of HIV--1--PR. "Hot" residues, playing a key role in the stability of the protein, are defined as those displaying the fastest modes. In this way regions 22--32, 74--78 and 84--91 are identified as those forming the folding core of the protein. These regions match with those displaying low experimental Debye-Weller factors, that is low fluctuations in the crystallographic structure.

Calculation of $\varphi$--values by means of G\=o--model simulations performed by Levy and coworkers \cite{levy} have located a major transition state where only regions 27--35 and 79--87 are structured. The protein then reaches the native state, overcoming another minor transition state, and subsequently dimerizes into the biologically active structure.

Cecconi and coworkers \cite{cecconi} have calculated the stability temperatures associated with each contact of the protease, again making use of a G\=o--model. They find that key sites to the stability of partially folded states, that is those displaying the lowest stability temperatures, are 22, 29, 32, 76, 84 and 86.

\subsection{Dynamics of contact formation}
 
To directly investigate the folding mechanism of the HIV--1--PR monomer, use is made of a simplified model which has proven useful to account for the folding properties of a number of small, single--domain proteins \cite{ludo}. The model pictures each amino acid as a spherical bead, making inextensible links with the rest of the chain. Each pair of amino acid interacts through a modified G\=o--contact potential \cite{go} 
\begin{equation}
U(\{r_i\})=\sum_{i<j}B_{ij}\Delta(|r_i-r_j|),
\end{equation}
where $r_i$ is the coordinate of the $i$th amino acid, $\Delta(|r_i-r_j|)$ is a contact function which assumes the value $1$ if the $i$th and $j$th residues are closer than $6.3\AA$ and zero otherwise, $B_{ij}$ being the energy matrix elements defined in Sect. 2b. Moreover, each amino acid displays a hard core of $3.8\AA$.

\begin{figure}
\centerline{\psfig{file=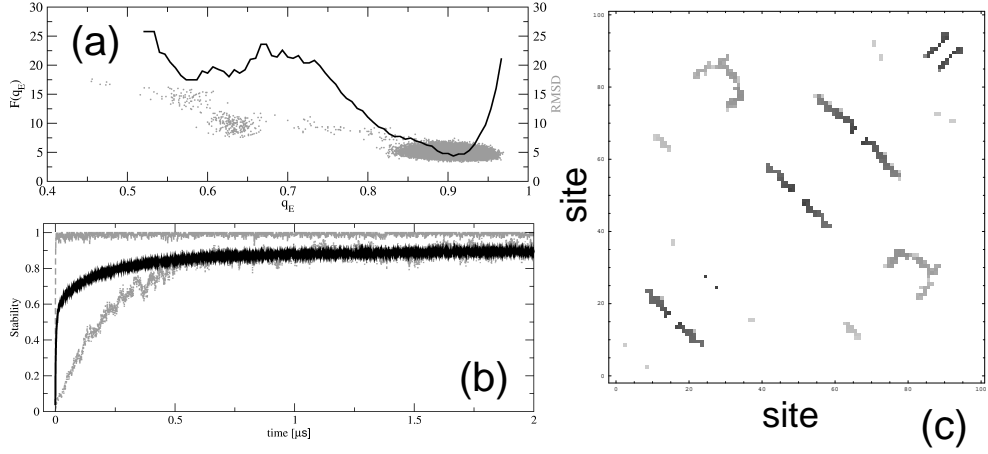,width=13cm}}
\caption{a) The free energy of HIV--1--PR monomer as a function of the relative energy (order parameter) $q_E$ (black curve) and the values of RMSD sampled during the simulation (gray dots) at $T=300K$; b) The formation probability of the native conformation (solid curve), the probability $p_{87-90}(t)$ (dashed gray curve) and the probability $p_{31-89}(t)$ (dotted gray curve), displayed as a function of time; c) Contact map of the HIV--1--PR, the different gray levels being associated with to different values of the average stabilization time. Darker symbols are formed in $10^{-10}$s while lighter symbols in $10^{-7}$s. The dynamics is simulated by mean of a dynamical Monte Carlo algorithm and the discrete time step is expressed in terms of time units by setting the simulated diffusion constant of the protein equal to $10^{-7} m/s^2$.}
\label{fig2}
\end{figure}

Equilibrium Monte Carlo simulations of the HIV--1--PR monomer provide free energy landscapes as a function of the fraction $q_E$ of native energy of the protein. The free energy at $T=300K$, shown in Fig. \ref{fig2}(a) with a solid curve, displays two major wells, corresponding to the unfolded and the native states. The grey beads mark conformations met during the sampling, and their position along the y--axis indicate the associated RMSD (cf. scale on the left y-axis). Note that the native well is quite broad, ranging from $q_E=0.75$ to $q_E=0.95$ and from an RMSD of 3 to 7 $\AA$. Also the free energy barrier to the unfolded state is rather small, being about 18 kJ/mol ($\approx 7\; kT$). These data indicate a two--state transition characterized by a weak degree of collectivity. To be noted that in the present calculations we only considered the monomeric form of the protein. Dimerization can increase the stability and the collectivity of the system.

Making use of the same model it is possible to run dynamical simulations, starting from random conformation and following the folding to the native state. The overall dynamics can be followed through the plot of $[q_E](t)$, that is the fractional native energy as a function of time, averaged over 100 independent runs. The result is displayed as a solid curve in Fig. \ref{fig2}(b), indicating an exponential process of characteristic time $\tau=2.9\cdot 10^{-7}$s, consistent with the two--well picture. The model provides also information about the formation of each contact, through the probability $p_{ij}(t)$ that the contact between residues $i$ and $j$ is formed at time $t$. A number of native contacts are stabilized early (sub--nanosecond time scale) following an exponential dynamics. This is the case for contacts belonging to $S_2$ (25--28) and in particular to $S_8$ (e.g. contacts 86--89, 87--90, 89--92, 90--93). Contacts between residues which are far along the chain are formed later, following a non--exponential dynamics, which indicate that their formation is dependent on some other event. The earliest involve the interaction between fragments $S_5-S_7$ and $S_2-S_8$ which take place after an average time of few and of tens of ns, respectively. As an example we display in Fig. \ref{fig2}(b) the formation probability of the contact 31-89 as a function of time.

In Fig. \ref{fig2}(c) is summarized the hierarchy of formation of native contacts of HIV--1--PR, the different gray levels corresponding to different time scales, while in Table \ref{table2} are listed the parameters associated to the selected contacts. The picture that emerges is that local contacts within fragments 83--93, between residues 25--28 and in the beta--hairpin 42--58 form first. Then the beta--turns 15--18 and 65--70, again built of local residues. The next event is the assembly of the nucleus involving fragments 22--34 and 83--93, which is further stabilized by the contribution of the strongly--interacting bend 77--83. Finally, the rest of the residues come to place.

Summing up, the model suggests that LES are built of residues which lie in the regions 83--93 and 22--34. This essentially agrees with the indirect indications provided by the studies of Levy and coworkers \cite{levy}, which indicate the formation of regions  27--35 and 79--87 as determinant for the folding of the protein. The stabilization core \cite{kolya} is somewhat larger (cf. Sect. \ref{sect_energy}), involving, besides residues  84--93 and 22--34, also the regions 56--66 and 74--78.

From the above discussion one can conclude that peptides p--S$_2$ and p--S$_8$ are the most likely candidates as folding inhibitors of the HIV--1--PR monomer. Because $S_8$ is well structured (it builds 10 internal contacts and an $\alpha$-helix turn), while $S_2$ has little internal structure, we shall select p--S$_8$ as the best candidate for a non--conventional inhibitor. It is expected that peptides p--S$_8$ will efficiently attach to the $S_2$--LES, thus blocking the formation of the folding nucleus or, in the case the protein is already folded, by profiting favourable fluctuations around the native conformation, to come into contact with the $S_2$--LES thus blocking the return of the monomer to the original native conformation. 

\section{Inhibition of the folding of the HIV--1--PR}

Within the framework of the G\=o model discussed above, Monte Carlo simulations were carried out for a system composed of a single 99mer and a number (from 1 to 5) of p-S$_8$ peptides, placed inside a box of side of 100 $\AA$ with periodic boundary conditions.

\begin{figure}
\centerline{\psfig{file=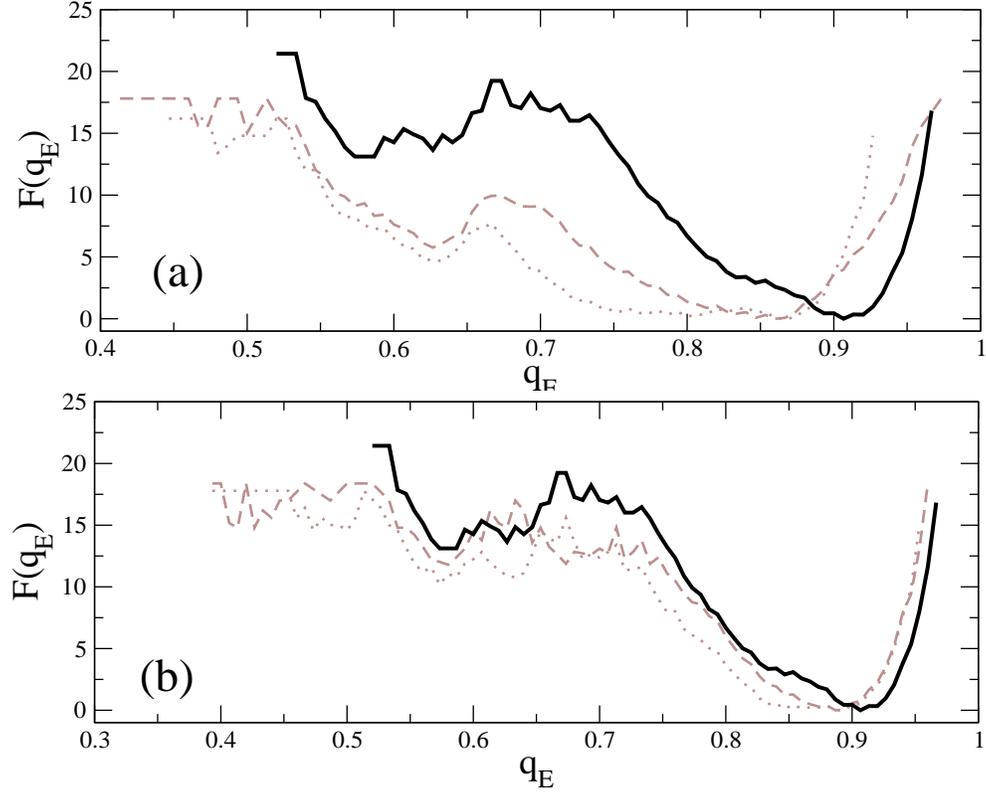,width=13cm}}
\caption{The free energy profile of the HIV--1--PR monomer (solid curve) and of the system composed of the monomer plus 3 (dashed) and 5 (dotted) p--S$_8$ peptides, as a function of the relative native energy $q_E$. The temperature at which the simulations were carried out is 300 K. b) The free energy profile of the monomer alone (solid curve) and of the system composed of 3 peptides corresponding to the fragments 9--19 (dotted curve) and 61--70 (dashed curve).}
\label{fig3a}
\end{figure}

In Fig. \ref{fig3a}(a) the resulting free energy of the protein as a function of $q_E$ is displayed. 
The presence of p--S$_8$ peptides has a number of effects: a) it broadens the minimum associated with the native conformation, b) it lowers by $\approx 10 kJ/mol$ the free energy of the minimum associated with the unfolded state, c) it lowers by about the same amount the free energy of the maximum of the barrier separating the two minima (transition state). These effects are associated with a decrease of the relative population  of the native state  from 0.98 to 0.78 in the case of 3 peptides and to 0.52  in the case of 5 peptides. Because the above results were obtained from calculations which describe the system at equilibrium they apply equally well  to situations in which one starts with the monomer in the state N or in state U.

To test the validity of these results, we have repeated the simulations using peptides corresponding to fragments of the protease monomers different from the LES, but having a similar number of amino acids. It is found that e.g.,  fragments 9--19  and 61--70 do not have any significant effect on the free energy of the native state (cf. Fig. \ref{fig3a}(b)). This result testifies to the fact that the denaturing effect observed in the results shown in Fig \ref{fig3a}(a) is due not to some unspecific interaction between the peptides and the protein, but to the very choice of LES--mimicking peptides.

\begin{figure}
\centerline{\psfig{file=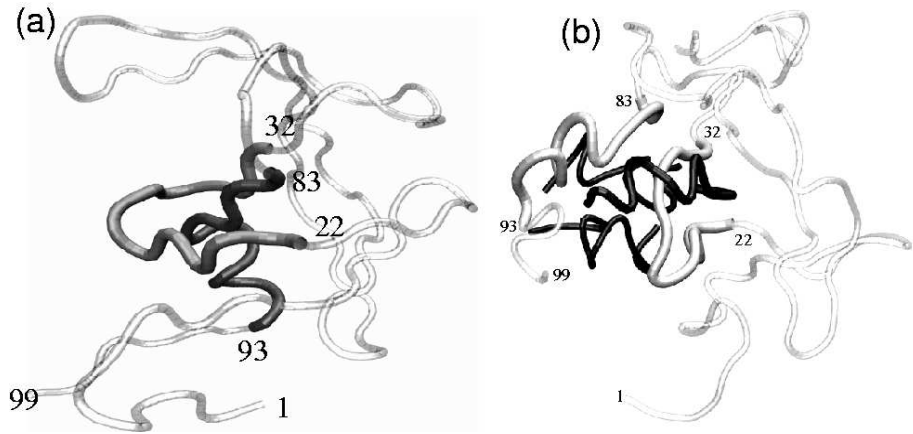,width=13cm}}
\caption{a) Native conformation of each monomer which build the HIV--1--PR dimer, based on the NMR conformation of ref. \protect\cite{native}. b) Snapshot of the sampling of the system composed of the protease and 3 p--S$_8$ peptides. The protein has been drawn in light grey and the S$_2$ and S$_8$ LES explicitely marked in dark grey. The p--S$_8$ peptides are shown in black. The ratio of contacts forming beta structure is $56\%$ for the native conformation shown in (c), and decreases to $43\%$ with 3 peptides (cf. (d)) and to $29\%$ with 5 peptides.}
\label{fig3b}
\end{figure}

To further clarify the mechanism which is at the basis of the results shown in Fig. \ref{fig3a}(a), we display in Fig. \ref{fig3b}(b) a snapshot of the simulation carried out with 3 peptides. It is seen that one of the p--S$_8$ peptides has built (strong) contacts stabilizing the folding nucleus of the protein. Consequently, the state corresponding to the correctly folded monomer (Fig. \ref{fig3b}(a)) and to the state in which the peptide binds to the partially folded protein have similar energies but very different free energies. In fact, the main difference between the two conformations displayed in Fig. \ref{fig3b}(a) and \ref{fig3b}(b) is entropic: while the partially folded protein has a much larger entropy as compared to the folded protein (which essentially is in a single conformation), the complexed state is penalized by the entropy decrease associated with the binding of a single peptide, decrease which can be reduced at will by increasing the number of peptides.

Consequently, whatever the folding entropy cost of binding a p--S$_8$ is, there exists a concentration of peptides which makes the complexed, partially--unfolded state favourable. In the case of the present simulations, a concentration of peptides equal to few times that of the protein seems to be adequate to consistently destabilize the protease.

We have also analyzed the effect pointlike mutations \footnote{We implement mutations by changing the interaction $B_{ij}$ of the mutated site $i$ with all its neighbour sites $j$ from the original (negative) value to $5\;kJ/mol$.}on the HIV--1--PR have on the inhibitng ability of p--S$_8$ peptides. Mutations on sites not belonging to LES have little or no effect on the stability of the protease, but at the same time do not affect the inhibitory action of p--S$_8$. For example, mutations  on site 73 cause the protease to decrease its stability at 300 K from $98\%$ (observed for the "wild--type" sequence) to $97\%$, the effect of 3 p--S$_8$ peptides leading to a decrease of this stability from $97\%$ to $32\%$ (to be compared with the $98\%$ to $52\%$ decrease observed in the case of the "wild--type" sequence). Mutations in sites belonging to the LES of the HIV--1--PR, on the other hand, cause destabilization of the native state of the protease, shifting the equilibrium distribution of the order parameter $q_E$ towards lower values. For example, mutation in site 34 causes the stability of the protease to diminish to $64\%$ (from the value of $98\%$).

\section{Spectrophotometric measurements}

Following the theoretical predictions discussed above, we have investigated the inhibitory properties of the peptide of sequence NIIGRNLLTQI (p--$S_8\equiv$ 83--93) by means of a spectrophotometric assay as described in ref. \cite{experiment} and making use of ultraviolet circular dichroism \cite{k2d}.

\subsection{Materials and methods}

Recombinant HIV-1-Protease, expressed in E. Coli (Sigma Cat. no. P7338) \cite{louis,roesel} contained five mutations to restrict autoproteolysis (Q7K, L33I, L36I) and to restrict cysteine thiol oxidation (C67A and C95A). The enzyme was stored at (-70C as solution with concentration $25\mu g/63\mu L$ in dilute HCl, (pH=1.6) ). A chromogenic substrate for HIV-1-PR (Sigma Cat. no. H5535) (with sequence Arg-Val-Nle-Phe(NO$_2$)-Glu-Ala-Nle-NH$_2$) was obtained as a 1 mg desiccate,  diluted with 0.1 ml of DMSO, and stored at -20C.  Protease assisted cleavage between the Nle and the Phe(NO$_2$) residues of substrate entails a blue-shift of the absorption maximum (277 nm to 272 nm). It can be adequately monitored observing the continuous decrease of absorbance at 310 nm \cite{experiment}. A regression of the absorbance at 300 nm against substrate concentration allows to estimate the extinction coefficient of the whole substrate ($S=6300\pm 1600\; (mol\;cm)^{-1}$). Moreover, the absorbance at 300 nm after complete peptidolysis allows to determine a differential extinction coefficent ($=1500\pm 250\;(mol\;cm)^{-1}$) between the whole substrate and the cleaved products. This compares well with a difference of extinction coefficient at 310 nm between the cleaved and the complete substrate of $1200\pm 100\;(mol\;cm)^{-1}$, reported in ref. \cite{experiment}.
    
Inhibitor peptide (peptide p--S$_8$) from the primary sequence of the HIV--1--PR wild type (PDB code 1BVG) were synthesized by solid phase synthesis (Primm, San Raffaele Biomedical Science Park, Milan) with acetyl and amide as terminal protection group  and was estimated to be $>95\%$ pure by analytical HPLC after purification. After that 1 mg of inhibitor peptide was dissolved in 100 $\mu$l of DMSO, 4 $\mu$l of this solution were then diluted with 16 $\mu$l of DMSO and 180 $\mu$l of the buffer used for assay. The obtained solution (150 $\mu$M of peptide p--S$_8$) was used for the experiments. 

The buffer was prepared by adding 0.8 mM NaCl, 1 mM EDTA and 1 mM dithiothreitol to a 20 mM phosphate buffer (pH 6).
Other peptides displaying a sequence equal to that of selected fragments of the protease, but not belonging to any LES, as well as another peptide displaying a random sequence, were also used in the same way as peptide p--S$_8$. The sequences of these control peptides are QILIEICGHK and PLVTIKIGGQL corresponding to fragments 61-70 and 9-19, respectively, of the HIV-1-PR sequence, as well as the LSQETFDLWKLLPEN sequence, which is not related in any way to the protease. We will call these peptides K1, K2 and K3, respectively. To be noted that peptide K2 is rather hydrophobic and only $>70\%$ purity could be achieved.

Each measure was performed recording the absorbance at 300 nm (4.13 eV) in a standard UV-vis spectrophotometer (Jasco V-560). The sample had a total volume of 70 $\mu L$ in Spectrosil Far UV Quartz (170-2700 nm) cuvettes (3.3 mm optical path). The sample in the cuvette was exposed to a constant temperature ($37\pm 0.05C$) provided by continuous circulation of water from a waterbath to the cell holder via a circulation pump.

After proper thermal stabilization of the substrate dissolved in the buffer (pH 6), the absorbance at 300 nm was recorded. We performed 6 measures spanning a range of substrate concentrations from 50 $\mu M$ to 250 $\mu M$. The reaction was initiated by adding 2.78 $\mu g$ of enzyme and recording the asbsorbance decrease at 300 nm for 1200 sec. In all the reported measures the enzyme has thus a nominal concentration of 1.8 $\mu M$. Since, to our knowledge, no accurate measure has been published of the dimerization constant of HIV--1--PR at pH=6.0, it is difficult to estimate precisely the active site concentration in our assay. It should however be noted that the active site concentration only affects the estimation of $k_{cat}$, and has no influence over the enzyme-inhibitor dissociation constants.

The assessment of the enzyme activity in presence of the inhibitor peptides followed a similar routine: the enzyme and the peptide (whose concentration are in a ratio 1:3) were incubated together for approximately 1 min and then added to the substrate. In all the reported measures the final concentration of inhibitor was $[I_0]=5.4\;\mu M$. The asbsorbance decrease at 300 nm was recorded for 1200 sec.
  
Ultraviolet CD spectra were recorded on a Jasco J-810 spectropolarimeter in nitrogen atmosphere at room temperature using 0.1 cm path-length quartz cell. Each spectrum was recorded between 260-200 nm. The data were collected at a rate of 10 nm/min with a wavelength step of 0.2 nm and a time constant of 2 s. The spectra were corrected with respect to the baseline and normalized to the aminoacidic concentration. The protein and the peptide were dissolved in a 20 mM phosphate buffer with 0.8 M NaCl at the same concentration used for the activity assays. The CD spectra were analyzed in terms of contribution of secondary structure elements \cite{k2d} using the {\it K2D} method based on comparison with CD spectra of proteins and peptides with known secondary structure.

\subsection{Results}

\begin{figure}
\centerline{\psfig{file=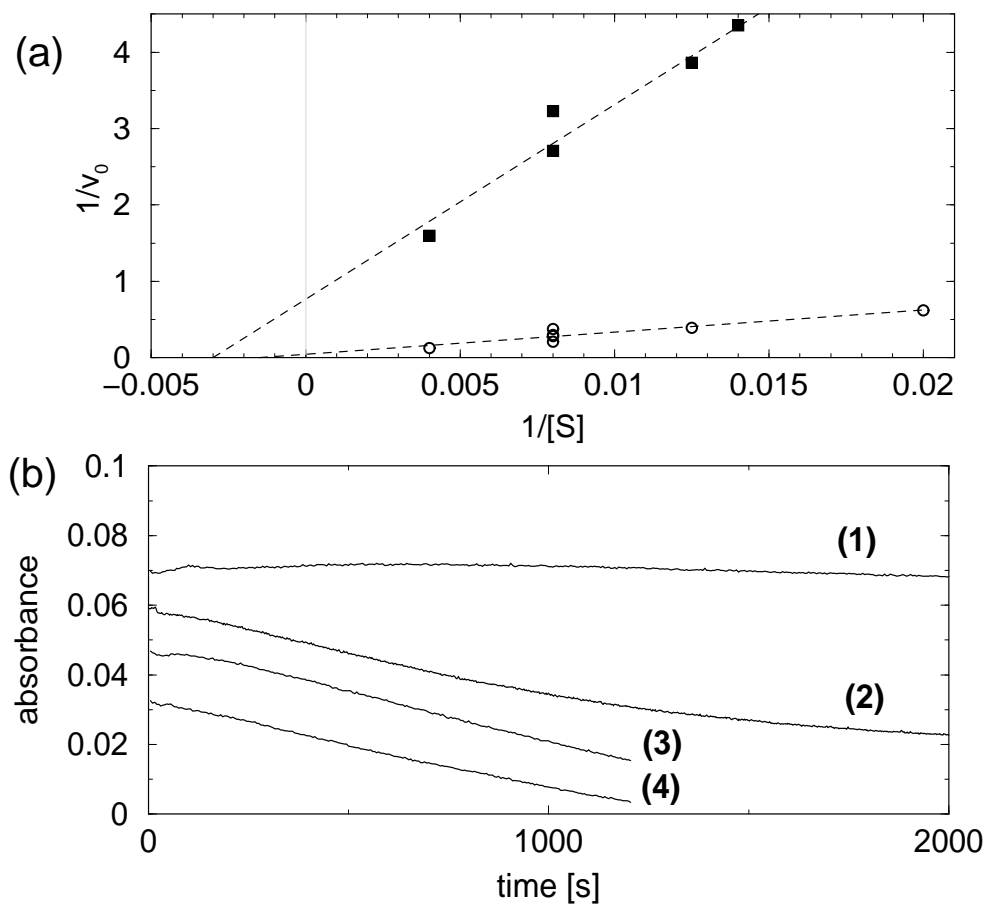,width=13cm}}
\caption{a) The Lineweaver-Burk plot associated with the protease (empty circles) and the protease complexed with the inhibitor p--S$_8$ (filled squares). The dashed lines are the linear fits to the experimental points. The initial velocities $v_0$ are expressed in terms of absorbance/s, while the substrate concentration $[S]$ is in mM. b) The enzymatic kinetics of the inhibited protease (1), of the protease alone (2), of the protease together with control peptides K1 (3) and K2 (4) , measured as change in absorbance of the chromogenic substrate as a function of time. All the curves have been measured at $[S_0] = 125 mM$, and have been shifted along the y-axis in order to be easily inspected.}
\label{fig4}
\end{figure}

In order to provide a quantitative measure of the enzymatic activity of HIV-1-PR and the inhibitory mechanism of  the designed peptides, we have performed a Michaelis-Menten analysis of the data reported in Fig. \ref{fig4}(b), measuring the initial reaction velocity $v_0$ at different substrate concentrations (cf. Fig. \ref{fig4}(a)). We obtain an estimate of the dissociation constant $K_M$ for the uninhibited-enzyme/substrate complex (curve (2) of Fig. \ref{fig4}(b)) of $670\pm 860\;\mu M$.  The maximum velocity of the reaction $v_{max}$ results $0.40\pm 0.52\; \mu M/s$, corresponding to a catalitic rate $k_{cat} = 0.25\pm 0.28 s^{-1}$. Note that the order of magnitude of these results compares well with those reported for the same substrate in ref. \cite{experiment}, that is $K_M=280\pm 100\mu M$ and $k_{cat} =7.3 \mu M/s$.
The apparent dissociation constant $K'_M$ of the inhibited reaction is $332\pm 191 \mu M$, the maximum velocity being $v_{max}=0.026\pm 0.011\mu M/s$ (cf. curve (1) of Fig. \ref{fig4}(b)). The fact that the value $v_{max}$ changes in presence of the inhibitor while that of $K'_M$ remains within  the experimental error of $K_M$ suggests that the inhibition is compatible with a non-competitive scenario \cite{kinetics}, in which the inhibitor can bind to the free enzyme with a dissociation constant $K_I$ or to the substrate-enzyme complex with a dissociation constant $K'_I$ not significantly different from $K_I$ . The results shown in Fig. \ref{fig4}(a) lead in fact to the values $K_I =380\pm 810 nM$ and $K'_I =180\pm 234 nM$. The caveats and the physical meaning of the interpretation of Fig. \ref{fig4}(a) in terms of a classical mixed models will be examined in the next section.

We note that the large errors in the estimate of the kinetic parameters are due to the small number of points recorded. An alternative method to determine the value of these parameters involves the fitting of the data with the analytical solution of the Michaelis-Menten kinetic \cite{schnell}, solution which can be written in terms of the reduced concentation $S'=S/K_M$ as: $[S'](t) = W([S'_0] exp(-k t + [S'_0]))$, $W$ being Euler's Omega function and $k=v_{max}/K_M$.  

Using a nonlinear fit to the data with the above expression one can obtain a more robust estimate of both $K_M$ and $v_{max}$ using the complete time-dependence of the concentration of free enzyme, expressed as a function of the measured absorbance as $[S](t) = 1/[A(t) - A(\infty)]/\delta\epsilon$. In particular, using this method on curve (2) of Fig. \ref{fig4}(b), we have been able to determine the kinetic values of the protease--substrate reaction obtaining $K_M=656\pm 320 \mu M$ and $k_{cat}=0.49\pm 0.08\;s^{-1}$. These values characterize more accurately the enzymatic process than the values obtained from the analysis reported in Fig. \ref{fig4}(a). 

We have repeated similar measures using this time the control peptides K1, K2 and K3 instead of peptide p--S$_8$, and found no appreciable variation in the kinetic parameters with respect to the uninhibited case (cf. curve (3) and (4) of Fig. \ref{fig4}(b)). In the presence of peptide K1 or K2 the reaction displayed initial velocities $v_0=0.051 \mu M/s$ and $v_0=0.049 \mu M/s$, respectively, essentially identical to the value of $v_0$ obtained from the global fit to the curve (2) of Fig. \ref{fig4}(b).

\begin{figure}
\centerline{\psfig{file=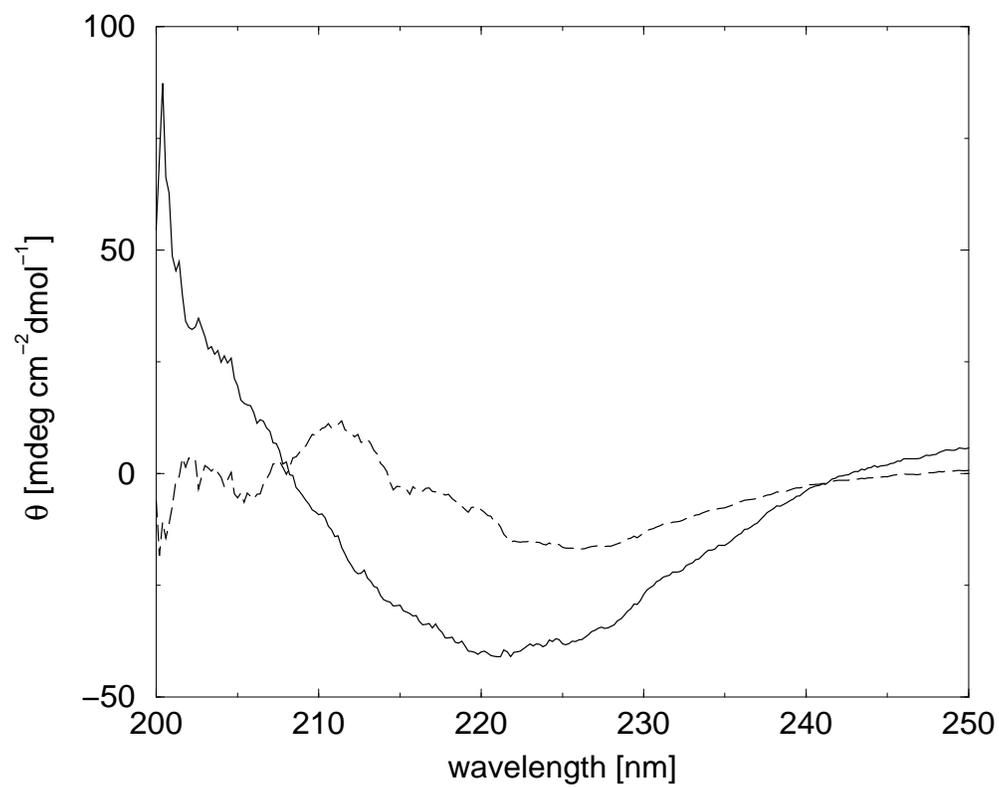,width=13cm,angle=-90}}
\caption{The circular dichroism spectrum of the protease (solid curve) and of the solution composed of the protease and p--S$_8$ peptide (dashed curve) in the ratio 1:3.}
\label{fig5}
\end{figure}

The conformational properties of the inhibited protease have been investigated by means of UV circular dichroism experiments. The CD spectrum of the protease (cf. Fig. \ref{fig5}) under the same conditions used for the activity assay indicate a beta-sheet content of  79\%, consistent with the beta character of the native conformation \cite{native}. The CD spectrum of the solution of protease plus p--S$_8$ inhibitor, at the same concentrations and under the same conditions as those of the activity assay, shows a loss of beta--structure  (to a beta--sheet content value of 45\%), indicating that the protein is, to a large extent, in a non-folded conformation. These numbers compare well with those predicted by the model calculations (cf. caption to Fig. \ref{fig3b}).

\subsection{Discussion}

The results described above show that the p--S$_8$ peptide inhibits the activity of the HIV--1--PR, interefering with the native character of its equilibrium state. Although a Lineweaver-Burk plot like that displayed in \ref{fig4}(a) is usually interpreted in terms of a non-competitive kind of inhibition, it can also describe the case of a irreversible binding of the inhibitor to the enzyme \cite{kinetics}. In fact, in a reaction of the kind (where $I\equiv$p--S$_8$) 
\begin{equation}	
\begin{array}{lllll}
E+S & \overset{k_{+S}}{\underset{k_{-S}}{\rightleftarrows}} & ES & \overset{k_{cat}}{\rightarrow} & P \\
+ & & & & \\
\multicolumn{3}{l}{I \overset{k_{+I}}{\underset{k_{-I}}{\rightleftarrows}} EI} & & \\
\end{array}
\end{equation}
where the rates $k_{+I}$ and $k_{-I}$ are much smaller than  $k_{+S}$ and $k_{-S}$, one can consider the concentration $[EI]$ as constant on a time scale smaller than $(k_{-I})^{-1}$, and given by $[EI] = (k_{-I} /k_{+I}) ([E]_0-[EI])[I]_0$. In this context, irreversibility means that on the time scale relevant for the experiment the dissociation of EI into E and I is negligible. The concentration of enzyme available to the equilibrium with substrate is then, on this time scale, $[E]_0-[EI]$. This leads to a Michaelis-Menten equation identical to that associated with non-competitive inhibition, of the form
\begin{equation}
v_0=\frac{[S]_0[E]_0\left(1+\frac{[I]_0}{k_I}\right)^{-1}k_{cat}}{[S]+K_M}
\end{equation}
where we have set $k_i=k_{-I}/k_{+I}$ . Note that the difference between the irreversible scenario and the reversible competitive case is that, in the former, $[EI]$ is proportional to $[E0]$ while in the latter $[EI]$ is proportional to the free enzyme concentration $[E]$. Such irreversible scenario is not unexpected in the case of folding inhibitors, where the inhibitor binds tightly to the enzyme (as tight as the interactions which stabilize the protein), the unfolding rate being of the order of $10^{-2} s^{-1}$ \cite{szeltner}, giving rise to a value of $(k-I)^{-1}$ of the order of minutes.

\section{Conclusions}

The peptide p--S$_8$ ($\equiv$83N, 84I, 85I, 86G, 87R, 88N, 89L, 90L, 91T, 92Q, 93I) displaying a sequence identical to that of the S$_8$ LES of each of the two identical chains forming the HIV--1--PR homodimer is found to be a highly specific and likely quite efficient inhibitor ($K_I=380\pm 810\;nM$) of the folding of the 99mers, and thus of the whole enzyme. A remarkable property of this inhibitor is that it is unlikely that it would allow for escape mutants. In fact, the only mutations which will prevent p--S$_8$ from acting are expected to involve protein denaturation.

Obvious disadvantages of the present design are the length and the peptidic character of the inhibitor, because it is not clear how to prevent the degradation by cellular enzymes. Consequently, there are two clear tasks lying ahead. One, to investigate  whether the shortening of p--S$_8$, by leaving out some residues either at the beginning or at the end (or both), lead to peptides which still inhibit folding with similar specificity and effectiveness as p--S$_8$ does. The second is to develop molecules mimetic to p--S$_8$ or eventually to shorter peptides derived from p--S$_8$.

We conclude by suggesting that the strategy employed to design p--S$_8$, being universal, can be used to design inhibitors of the folding of other target proteins. 

The support and advice of R. Jennings, F. Garlaschi, G. Zucchelli and E. Ragg is gratefully acknowledged. We wish also to thank G. Carrea and G. Colombo for discussions.


\begin{table}
\begin{tabular}{|c|c|c|c|c|c|c|c|c|}
\hline
 & S$_1$ & S$_6$ & S$_8$ & S$_2$ &S$_7$ & S$_5$ & S$_3$ &  S$_4$ \\\hline
S$_1$ & -151.2 & -41.5 & -7.3 & -8.0 & - & -52.2 & -4.1 & - \\\hline
S$_6$ & -41.5 & -175.9 & -47.9 & -5.4 & -2.9 & - & - & -  \\\hline
S$_8$ & -7.3 & -47.9 & -201.4 & -110.5 & -7.4 & -20.1 & - & - \\\hline
S$_2$ & -8.0 & -5.4 & -110.5 & -115.9 & -57.5 & -4.4 & -4.8 & -2.0 \\\hline
S$_7$ & - & -2.9 & -7.4 & -57.5 & -42.0 & -92.8 &  -7.4 & - \\\hline
S$_5$ & -52.5 & - & -20.1 & -4.4 & -92.8 & -143.7 & -152.3 & -6.8 \\\hline
S$_3$ & -4.1 & - & - & -4.8 & -7.4 & -152.3 & -118.3 & -96.4 \\\hline
S$_4$ & - & - & - & -2.0 & - & -6.8 & -96.4 & -70.9 \\\hline
\end{tabular}
\caption{The all atom molecular dynamics energy map of the eight amino acid chain segments with which one can represent in an economic fashion each of the two HIV--1--PR monomers.}
\label{table1}
\end{table}

\begin{table}
\begin{tabular}{|c|c|c|c|}
\hline
contact & $B_{ij}$ & $p_\infty$ & $\tau$ \\\hline
25-28 & -5.4 & 0.77 & $2\cdot 10^{-10}$ \\
87-90 & -9.8 & 0.99 & $4\cdot 10^{-10}$ \\
31-89 & -6.1 & 0.92 & $2.6\cdot 10^{-7}$ \\
23-85 & -4.9 & 0.86 & $3.6\cdot 10^{-7}$\\
62-74 & -2.4 & 0.51 & $1.0\cdot 10^{-6}$ \\
12-66 & -1.5 & 0.51 & $1.2\cdot 10^{-6}$ \\
\hline
\end{tabular}
\caption{The dynamics of some native contacts of the protein. $B_{ij}$ is the interaction energy expressed in kJ/mol, $p_\infty$ is the asymptotic stability and $\tau$ is the average formation time of the contact in seconds. }
\label{table2}
\end{table}


\end{document}